\documentclass[aps,pra,showpacs,groupedaddress,superscriptaddress,twocolumn]{revtex4-1}
\usepackage{amssymb}
\usepackage{graphicx}
\usepackage{amsmath}
\usepackage{subfigure}
\usepackage{float}
\usepackage{multirow}
\usepackage{mathrsfs}

\newcommand{\bear}{\begin{eqnarray}}
\newcommand{\eear}{\end{eqnarray}}
\newcommand{\be}{\begin{equation}}
\newcommand{\ee}{\end{equation}}
\newcommand{\beqn}{\begin{eqnarray}}
\newcommand{\eeqn}{\end{eqnarray}}
\newcommand{\beqnn}{\begin{eqnarray*}}
\newcommand{\eeqnn}{\end{eqnarray*}}

\begin{document}

\title{Controlling properties of a hybrid Cooper pair box
interacting with a nanomechanical resonator in the presence of Kerr
nonlinearities and losses.}
\author{C.~Valverde}
\email{valverde@ueg.br}
\affiliation{C\^{a}mpus Henrique Santillo, Universidade
 Estadual de Goi\'{a}s - 75.132-903, An\'{a}polis, Goi\'as, Brazil} 
\affiliation{Universidade Paulista (UNIP) - 74.845-090, Goi\^ania, Goi\'as,
Brazil} 
\affiliation{Escola Superior de Neg\'{o}cios (ESUP) - 74.840-090,
Goi\^{a}nia, Goi\'as, Brazil}

\author{A. N. ~Castro}

\affiliation{Universidade Paulista (UNIP) - 74.845-090,
Goi\^ania, Goi\'as, Brazil}

\author{B.~Baseia}
\affiliation{Instituto de F\'{\i}sica, Universidade Federal de Goi\'as,
74001-970 Goi\^ania, GO, Brazil} 
\affiliation{Departamento de
F\'{\i}sica, Universidade Federal da Para\'{\i}ba - 58.051-970, Jo\~{a}o
Pessoa, Para\'{\i}ba, Brazil}

\date{\today }

\begin{abstract}
We consider the Jaynes-Cummings model describing the interaction of a Cooper
pair box (\textit{CPB}) and a nanoresonator (\textit{NR}) in the presence of
a Kerr medium and losses The evolution of the entropy of both subsystems and
the \textit{CPB} population inversion were calculated numerically. It is
found that these properties increase when the \textit{NR} frequency is
time-dependent, even in the presence of losses; the effect is very sensitive
to detuning and disappears in the resonant regime. The roles played by the
losses affecting the \textit{CPB} and the \textit{NR} are also compared.
\end{abstract}

\maketitle

\section{Introduction}

The implementation of the quantum computer is considered as a big challenge
in the recent years due to the failure of classical computers ability to
simulate and compute the large scale data. The interaction of the single
photon field with superconducting circuit to generate quantum bits is
studied experimentally \cite{15c,16c} and theoretically \cite{24c,25c,26c}.
Wallraff et al. have observed a strong\ coupling between the superconducting
two-level system and a single microwave photon which prove the generation of
qubits. Recently, the construction of the first quantum computer by Canadian
company (D-Wave) using the system of the superconducting qubits was shown.
This quantum computer succeeds to solve famous problem in mathematics, the
Ramsey numbers \cite{31c}.

The study of the interaction between a two-level atom and a radiation field
is very important in quantum mechanics, as its applications in laser physics
and quantum optics. This system has similarities with others, e.g., the
interaction between a Cooper pair box (\textit{CPB}) and a nanomechanical
resonator (\textit{NR}). There are many works in the literature dealing with
these systems \cite{c3,c4}, but only few of them treat these systems in the
situation where one of the frequencies \cite{15,15a}, or the amplitude \cite%
{16}, varies with time. In the \textit{CPB}-\textit{NR }system these
variations change the subsystems coupling and modify their dynamic
properties, e.g., showing\emph{\ }amplification of the excitation transition
rate in subsystems \cite{17}.

An important focus of quantum optics is concerned with the atom-field
coupled system. Inspired by various tests applied to this system, and its
limitations, the researchers have passed from the light domain to the
microwave domain of superconducting cavities coupled to Rydberg atoms, or to
the quantum electrodynamics circuit coupled to nanoresonators. After being
extended to broader context the system was used to investigate Landau--Zener
transitions \cite{nn3}; atomic physics and quantum optics \cite{nn4};
mechanisms for photon generation from quantum vacuum \cite{nn5}; quantum
simulation, challenges, and promises of fast-growing field \cite{nn6}.\ Here
we will assume the laboratory substituting the atom (field) by the \textit{%
CPB }(\textit{NR}) \cite{nn2}. This system have been analyzed in several
works, e.g.: quantum network \cite{nn7}; phonon blockade \cite{nn8};
squeezed states and entangled states \cite{nn9}; cooling mechanical
oscillators \cite{nn10}; Bell inequality violations \cite{nn11}.

In this report we will employ the Jaynes-Cummings model to treat a \textit{%
CPB}-\textit{NR}\ system with losses; a nonlinear Kerr medium is also added
to include influences of nonlinearities. We investigate the time evolution
of the \textit{CPB}\ population inversion ($I(t)$), as well as\ the
statistical properties of both subsystems. We consider the \textit{NR}\
initially in a coherent state and the \textit{CPB}\ in its excited state.
The influence caused by the time-dependent \textit{NR}\ frequency upon the
properties of both subsystems is studied. We consider the combined effects
of nonlinearity and losses upon the dynamics of the \textit{CPB}\ population
inversion and upon the \textit{NR}\ entropy, the latter being used as a
measure for the degree of entanglement. We compare the results obtained in
the resonant case ($\omega _{NR}=\omega _{CPB})$ with those obtained for
small detunings between the \textit{CPB}\ and \textit{NR}. The influences of
losses from the \textit{NR} and \textit{CPB} upon the mentioned properties
are also compared.

\section{The Hamiltonian System}

\begin{figure*}[h!tb]
\centering  
\includegraphics[width=20cm, height=8cm]{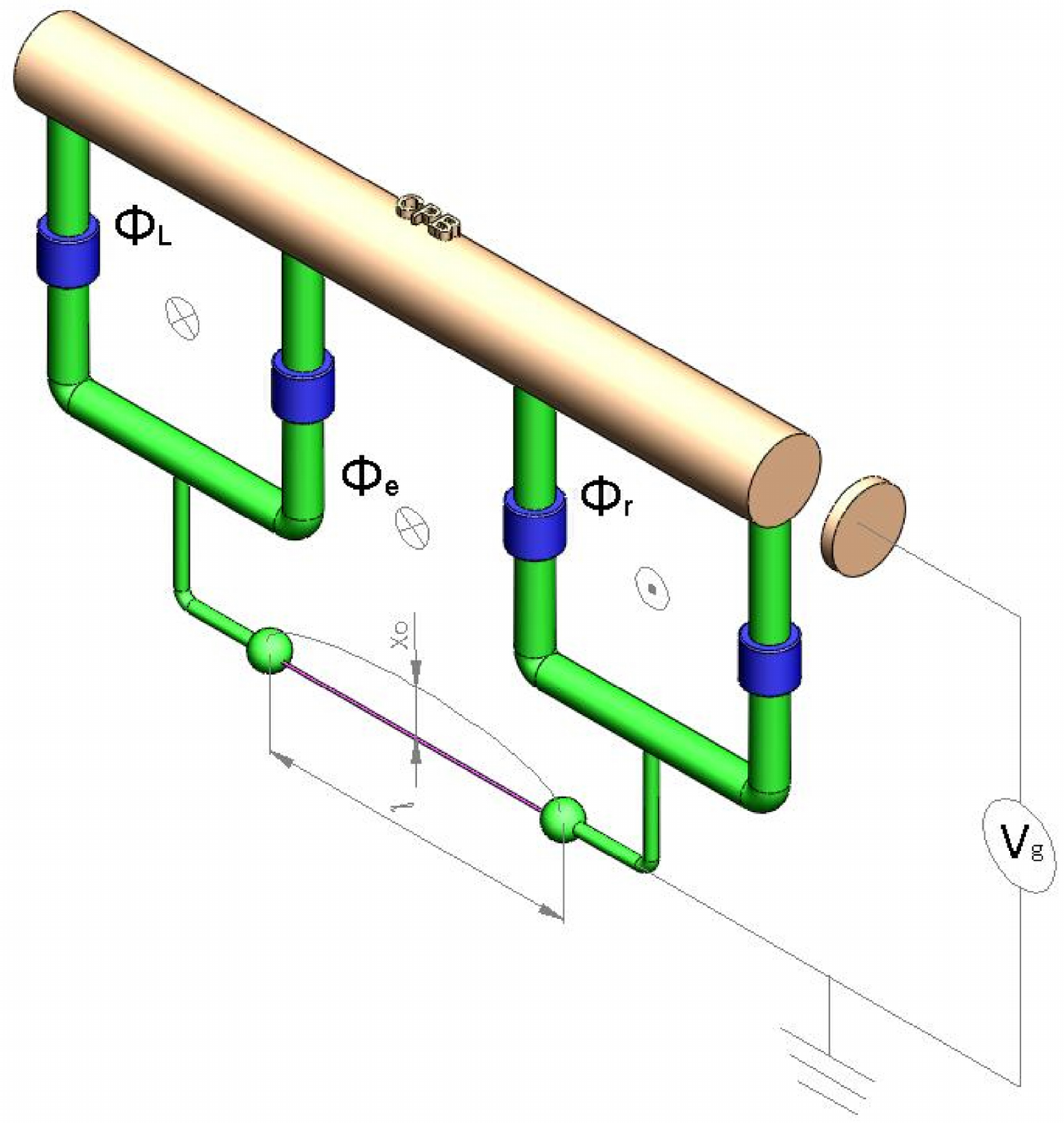}
\caption{\textit{Model for the CPB-NR coupling.}}
\label{cooper}
\end{figure*}

A superconductor \textit{CPB }charge qubit is adjusted\ to the input voltage 
$V_{g}$ of the system through a capacitor with an input capacitance $C_{1}$.
Observing the configuration shown in Fig. (\ref{cooper}) we see three loops:
a small loop in the left, another in the right, and a great loop in the
center. The control of the external parameters of the system can be
implemented via the input voltage $V_{g}$ and the three external fluxes $%
\Phi _{L},$ $\Phi _{r}$ and $\Phi _{e}$.\ The control of theses parameters
allows us to make the coupling between the \textit{CPB} and the \textit{NR}.
In this way one can induce small neighboring loops. The great loop contains
the \textit{NR} and its effective area in the center of the apparatus
changes as the \textit{NR} oscillates, which creates an external flux $\Phi
_{e}(t)$ that provides the coupling between the \textit{CPB} and the \textit{%
NR}. 

We will consider $\hslash =1$ and assume as identical the four Josephson
junctions, with the same Josephson energy $E_{J}^{0}$, the same being
assumed for the external fluxes $\Phi _{L}$ and $\Phi _{r}$, i.e., with same
magnitude, but opposite sign: $\Phi _{L}=-\Phi _{r}=\Phi _{x}$. In this way
we can write the Hamiltonian describing the entire system as 
\begin{widetext}
\begin{equation}
\hat{H}=\omega_{0} \hat{a}^{\dagger }\hat{a}+4E_{c}\left( N_{g}-\frac{1}{2}%
\right) \hat{\sigma}_{z}-4E_{J}^{0}\cos \left( \frac{\pi \Phi _{x}}{\Phi _{0}%
}\right) \cos \left( \frac{\pi \Phi _{e}}{\Phi _{0}}\right) \hat{\sigma}_{x},
\label{ac1}
\end{equation}
\end{widetext}where $\hat{a}^{\dagger }(\hat{a})$ is the creation
(annihilation) operator for the \textit{NR}\ excitations, corresponding to
the frequency $\omega $ and mass $m$; $E_{J}^{0}$ and $E_{c}$ are
respectively the energy of each Josephson junction and the charge energy of
a single electron; $C_{1}$ and $C_{J}^{0}$ stand for the input capacitance
and the capacitance of each Josephson tunnel, respectively; $\Phi _{0}=\frac{%
\pi }{e}$ is the quantum flux and $N_{g}=\frac{C_{1}V_{g}}{2e}$ is the
charge number in the input with the input voltage $V_{g}$. We have used the
Pauli matrices to describe our system operators, where the states $%
\left\vert 0\right\rangle $ and $\left\vert 1\right\rangle $ represent the
number of extra Cooper pairs in the superconducting island. So we have,\emph{%
\ }$\hat{\sigma}_{z}=\left\vert 1\rangle \langle 1\right\vert -\left\vert
0\rangle \langle 0\right\vert $ and $E_{c}=\frac{e^{2}}{C_{1}+4C_{J}^{0}}.$

The magnetic flux can be written as the sum of two terms, 
\begin{equation}
\Phi _{e}=\Phi _{b}+B\ell x\text{ },  \label{ac4}
\end{equation}%
where the term $\Phi _{b}$ is the induced flux, corresponding to the
equilibrium position of the \textit{NR} and the second term describes the
contribution due to the \textit{NR} vibration; $B$ represents the magnetic
field created in the loop and $\ell $ is the length of the\ \textit{NR}. We
write the displacement $\hat{x}$ as $\hat{x}=x_{0}(\hat{a}^{\dagger }+\hat{a}%
),$ where $x_{0}$ represents the \textit{NR} amplitude oscillation.
Substituting the Eq.(\ref{ac4}) in Eq.(\ref{ac1}) and controlling the flux $%
\Phi _{b}$ we can adjust $\cos \left( \frac{\pi \Phi _{b}}{\Phi _{0}}\right)
=0$ to obtain 
\begin{widetext}
\begin{equation}
\hat{H}=\omega_{0} \hat{a}^{\dagger }\hat{a}+4E_{c}\left( N_{g}-\frac{1}{2}%
\right) \hat{\sigma}_{z}-4E_{J}^{0}\cos \left( \frac{\pi \Phi _{x}}{\Phi _{0}%
}\right) \sin \left( \frac{\pi B\ell \hat{x}}{\Phi _{0}}\right) \hat{\sigma}%
_{x}  \label{ac8}
\end{equation}
\end{widetext}and making the approximation $\frac{\pi B\ell x}{\Phi _{0}}$ $%
<<1$ we find the Hamiltonian in the form, $\hat{H}=\omega _{0}\hat{a}%
^{\dagger }\hat{a}+\frac{1}{2}\omega _{c}\hat{\sigma}_{z}+\lambda _{0}(\hat{a%
}^{\dagger }+\hat{a})\hat{\sigma}_{x},$ where the constant coupling $\lambda
_{0}$ and the effective energy $\omega _{c}$ are given by $\lambda
_{0}=-4E_{J}^{0}\cos \left( \frac{\pi \Phi _{x}}{\Phi _{0}}\right) \left( 
\frac{\pi B\ell x_{0}}{\Phi _{0}}\right) ,$ $\omega _{c}=8E_{c}\left( N_{g}-%
\frac{1}{2}\right) .$ In the rotating wave approximation the above
Hamiltonian results in the form, 
\begin{equation}
\hat{H}_{eff}=\omega _{0}\hat{a}^{\dagger }\hat{a}+\frac{1}{2}\omega _{c}%
\hat{\sigma}_{z}+\lambda _{0}(\hat{\sigma}_{+}\hat{a}+\hat{a}^{\dagger }\hat{%
\sigma}_{-}).
\end{equation}

Next, we consider a more general scenario, substituting $\omega \rightarrow
\omega (t)=\omega _{0}+f\left( t\right) ,$ $\lambda _{0}\rightarrow \lambda
(t)=\lambda _{0}(1+f(t)/\omega _{0})^{1/2}$ and $\chi (t)=\chi
_{0}+\varepsilon f(t)$ \cite{c1,c2}. In addition we consider the presence of
the term $\kappa (t)$ standing for the time-dependent loss affecting the 
\textit{CPB}, the term $\delta (t)$ being the same for the \textit{NR}, and $%
\chi (t)$ is the response time of the Kerr medium. This extended and
somewhat realistic scenario requires the substitution of the Hamiltonian $%
\hat{H}_{eff}$ by the Hamiltonian $\hat{\mathscr{H}}$ given by

\begin{widetext}
\begin{equation}
\hat{\mathscr{H}}=\omega (t)\hat{a}^{\dagger }\hat{a}+\frac{1}{2}\omega _{c}(t)\hat{%
\sigma}_{z}+\lambda (t)\left( \hat{a}\hat{\sigma}_{+}+\hat{a}^{\dagger }\hat{%
\sigma}_{-}\right) +\chi (t)\hat{a}^{\dagger 2}\hat{a}^{2}-\frac{i}{2}\kappa
(t)\left\vert 1\rangle \langle 1\right\vert -\frac{i}{2}\delta (t)\hat{a}%
^{\dagger }\hat{a}.  \label{a1}
\end{equation}%
\end{widetext}

In the Eq. (\ref{a1}) the first (second) term describes the \textit{NR }(%
\textit{CPB}) subsystem, the third term\ represents the \textit{CPB-NR}
coupling, the fourth term stands for the time dependence of the Kerr medium,
the fifth and the sixth terms,\ $\kappa (t)$\ and $\delta (t),$ were
mentioned above. The operators $\sigma _{\pm }=\sigma _{x}\pm i\sigma _{y}\ $%
leads to $\hat{\sigma}_{+}=\left\vert 1\rangle \langle 0\right\vert $,$\ 
\hat{\sigma}_{-}=\left\vert 0\rangle \langle 1\right\vert $; $\kappa (t)$\
is the \textit{CPB} decay coefficient from the excited level $\left\vert
1\right\rangle $ to the fundamental state $|0\rangle $, $\delta (t)$ is the
same for the \textit{NR} and $\lambda (t)$ is the time-dependent coupling
between the \textit{CPB }and \textit{NR}. Actually, the inclusion of loss in
the system turns its treatment somewhat more realistic since dissipation is
ubiquitous in the real world; as consequence the Eq. (\ref{a1}) corresponds
to the evolution of a system described by a non-Hermitian Hamiltonian \cite%
{d1,d2,d3,d4,d5}. On the other hand, as the response time of the Kerr medium
is assumed so fast the medium follows the \textit{NR} adiabatically and the
third-order nonlinear susceptibility can be modulated by the \textit{NR}
frequency $\omega \left( t\right) $.

The wave function that describes our system can be written as,

\begin{equation}
\left\vert \Psi \left( t\right) \right\rangle =\sum_{n}^{\infty }\left[
C_{1,n}\left( t\right) \left\vert 1,n\right\rangle +C_{0,n}\left( t\right)
\left\vert 0,n\right\rangle \right] ,  \label{a2}
\end{equation}%
where $C_{1,n}\left( t\right) $ and $C_{0,n}\left( t\right) $ are
respectively the probability amplitudes of the states $\left\vert
1,n\right\rangle $ and $\left\vert 0,n\right\rangle $, namely, the \textit{%
CPB} in its excited state $\left\vert 1\right\rangle $ or ground state $%
\left\vert 0\right\rangle $ with elementary excitations $n$ in the \textit{NR%
}. As mentioned before, at $t=0$\ the system is decoupled, the \textit{CPB}
initially in its excited state $\left\vert 1\right\rangle $ and the \textit{%
NR} in a coherent state $\left\vert \alpha \right\rangle $, given by

\begin{equation}
\left\vert \alpha \right\rangle =\sum\limits_{n=0}^{\infty }F_{n}\left\vert
n\right\rangle .  \label{cb1}
\end{equation}

The total wave function in the initial state can be written as $\left\vert
\Psi (0)\right\rangle =\sum_{n}^{\infty }F_{n}\left\vert 1,n\right\rangle ,$
with the initial amplitudes $C_{0,n}(0)=0$ and $\sum_{n=0}^{\infty
}\left\vert C_{1,n}(0)\right\vert ^{2}=1.$ The Schr\"{o}dinger equation for
the present system, described by non-Hermitian and time-dependent
Hamiltonian in the Eq. (\ref{a1}) is,

\begin{equation}
\frac{d\left\vert \Psi \left( t\right) \right\rangle }{dt}=\hat{-i\mathscr{H}%
}\left\vert \Psi \left( t\right) \right\rangle .  \label{a3}
\end{equation}

We obtain the following set of coupled equations of motion for the amplitude
probabilities $C_{1,n}\left( t\right) $ and $C_{0,n+1}\left( t\right) $:

\begin{widetext}
\begin{equation}
\frac{\partial C_{1,n}\left( t\right) }{\partial t}=\left( -in\omega (t)-i%
\frac{\omega _{c}(t)}{2}-i\chi (t)(n^{2}-n)-\frac{1}{2}(\kappa (t)+n\delta
(t))\right) C_{1,n}\left( t\right) -i\lambda (t)\sqrt{(n+1)}C_{0,n+1}\left(
t\right) ,  \label{a9}
\end{equation}

\begin{equation}
\frac{\partial C_{0,n+1}\left( t\right) }{\partial t}=\left( -i(n+1)\omega
(t)+i\frac{\omega _{c}(t)}{2}-i\chi (t)(n^{2}+n)-\frac{1}{2}(n+1)\delta
(t)\right) C_{0,n+1}\left( t\right) -i\lambda (t)\sqrt{(n+1)}C_{1,n}\left(
t\right), \label{a10}
\end{equation}
\end{widetext}whose solutions allow us to calculate the entropy of the 
\textit{NR}\ subsystem and the population inversion of the \textit{CPB}.

\section{Time Evolution of the Population Inversion}

The present approach also allows us to investigate the \textit{CPB} dynamics
in a non perturbative way.\ A convenient way to characterize the response to
the \textit{NR} influence is given by the \textit{CPB} population inversion.
This parameter is defined as 
\begin{equation}
I(t)=\sum\limits_{n=0}^{\infty }\left[ \left\vert C_{1,n}(t)\right\vert
^{2}-\left\vert C_{0,n+1}(t)\right\vert ^{2}\right] .  \label{cc1}
\end{equation}

To calculate this property $I(t)$ we will assume the \textit{NR} frequency
varying with time as $\omega \left( t\right) =\omega _{0}+f(t)$. The third
order nonlinear susceptibility is modulated as $\chi (t)=\chi
_{0}+\varepsilon f(t)$ and we also assume the \textit{NR} initially in a
coherent state with the mean excitation number$\ \bar{n}=25$ and $\frac{%
\omega _{a}}{\lambda _{0}}=\frac{\omega _{0}}{\lambda _{0}}=20k$. We
consider the time evolution of the population inversion for different values
of the decay coeficients $\kappa (t)$ and $\delta (t)$.

\subsection{Resonant case: $f(t)=0$}

\begin{figure}[h!tb]
\centering
\subfigure(a){\includegraphics[height=6cm]{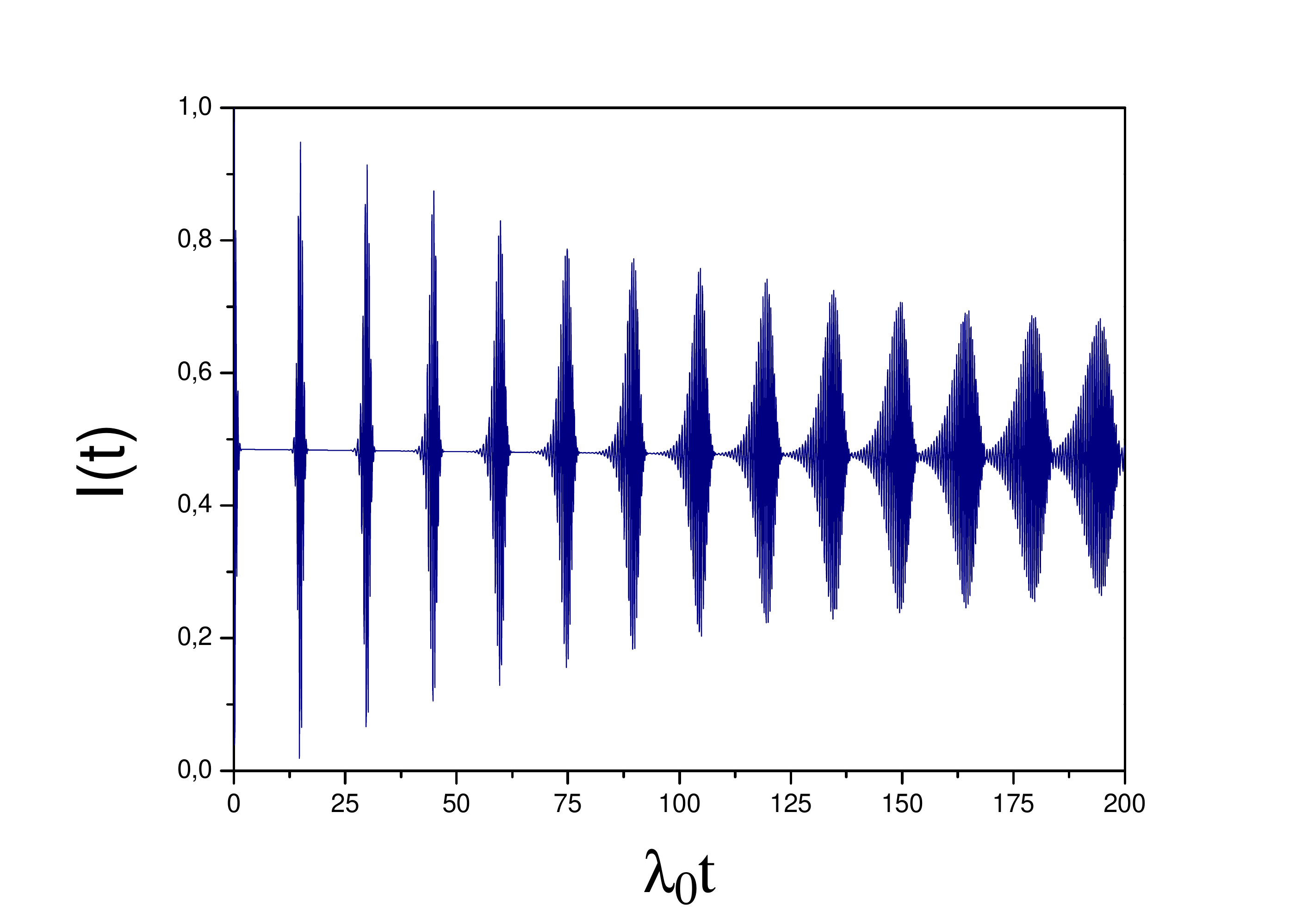} \label{fig7} } \quad 
\subfigure(b){\includegraphics[height=6cm]{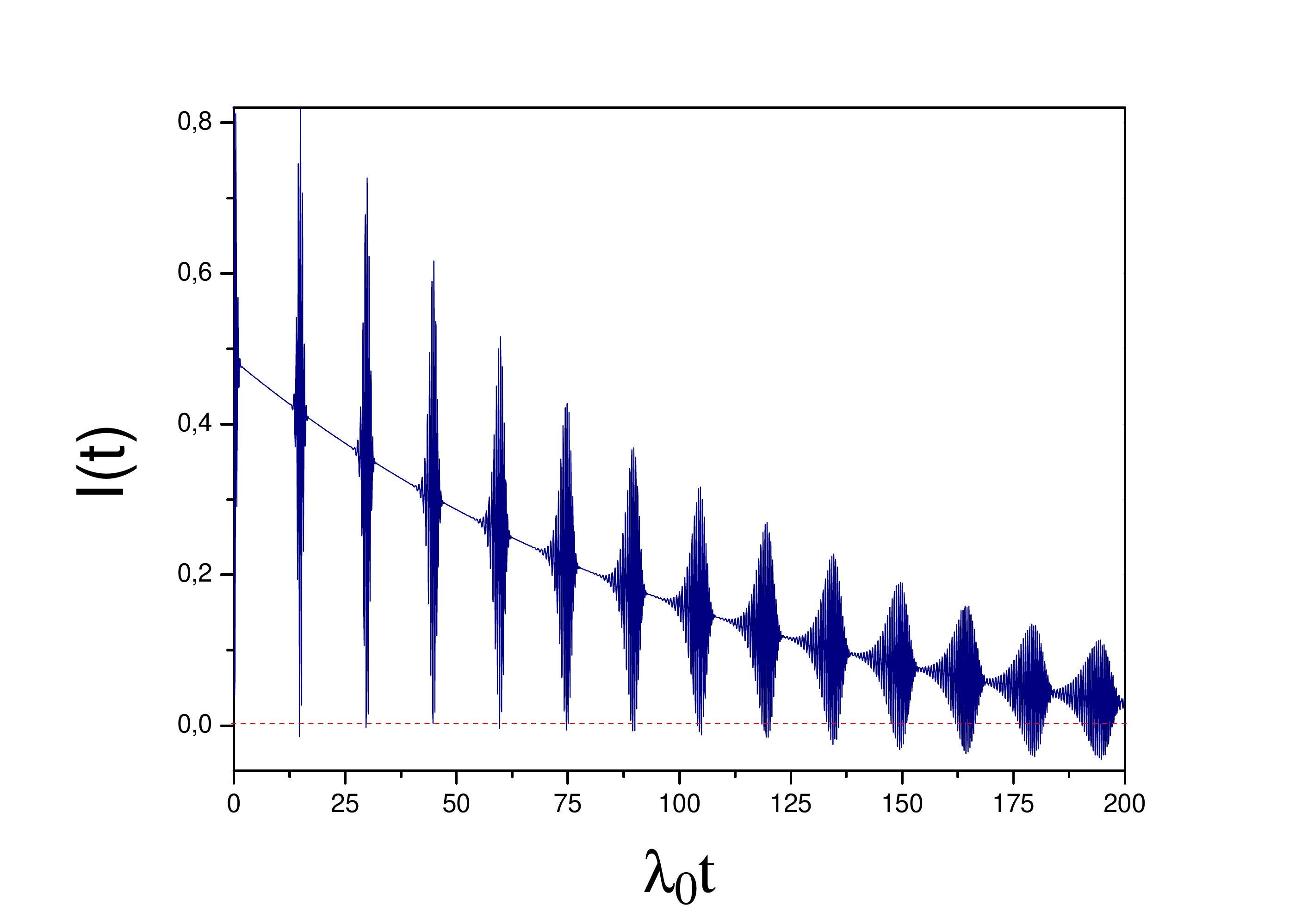} \label{fig8} } \quad 
\subfigure(c){\includegraphics[height=6cm]{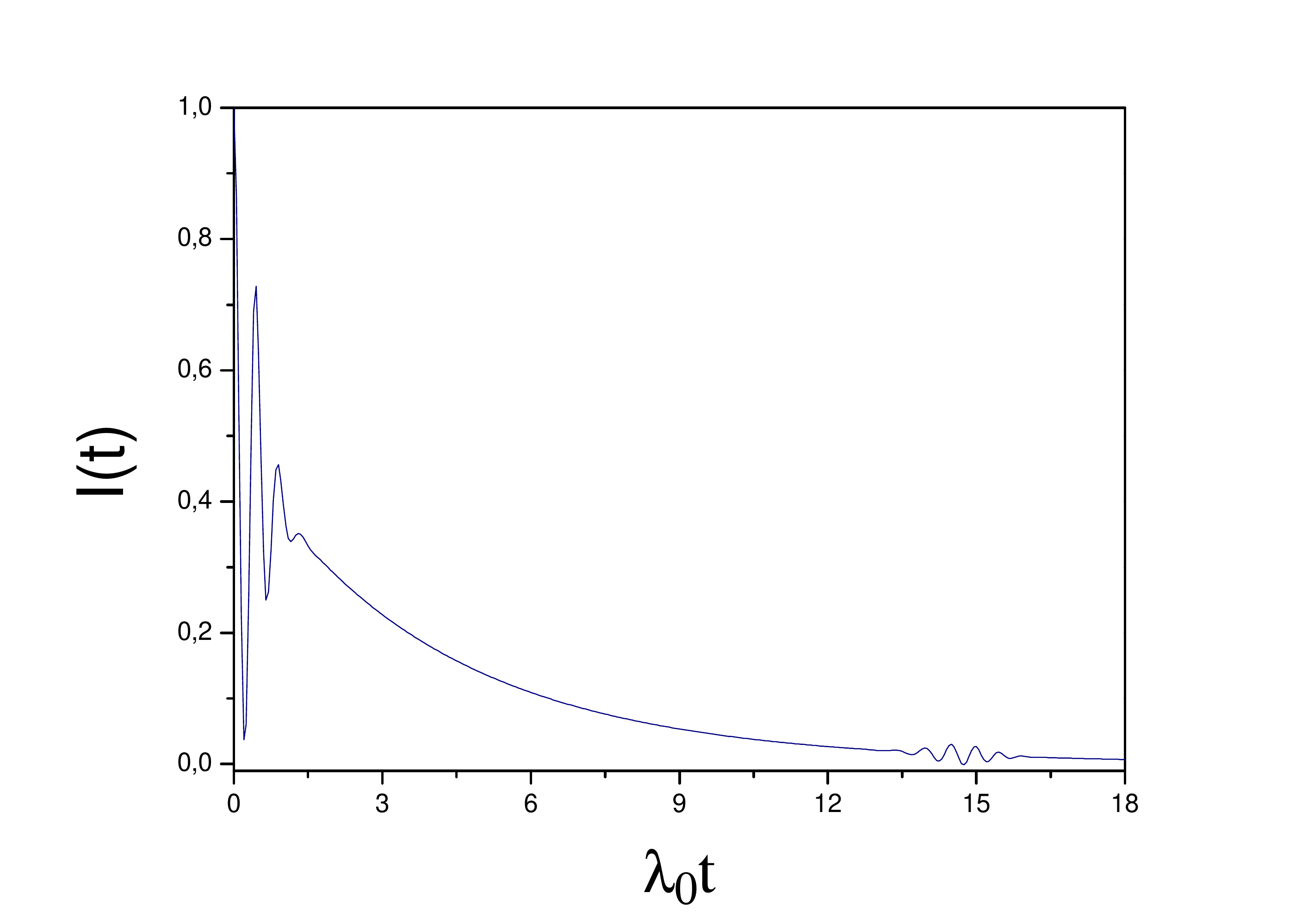} \label{fig9} } \quad 
\caption{Time evolution of the population inversion for different values of
the parameters $\protect\kappa (t)$ and $\protect\delta (t)$ for: $%
\left\langle n\right\rangle =25$, $\protect\omega _{0}/\protect\lambda _{0}=%
\protect\omega _{c}/\protect\lambda _{0}=20k$, $f(t)=0$; $\protect\chi _{0}/%
\protect\lambda _{0}=0.2;$ (a) $\protect\kappa /\protect\lambda _{0}=0.0$
and $\protect\delta /\protect\lambda _{0}=0.0$; (b) $\protect\kappa /\protect%
\lambda _{0}=0.01$ and $\protect\delta /\protect\lambda _{0}=0.0$; (c) $%
\protect\kappa /\protect\lambda _{0}=0.0$ and $\protect\delta /\protect%
\lambda _{0}=0.01.$}
\label{figInv1}
\end{figure}

If the model is restricted to the usual case, with the \textit{NR} frequency
constant, the Eqs. (\ref{a9} and \ref{a10}) can be solved numerically. In
Figure (\ref{figInv1} a) we plotted the inversion $I(t)$ as a function of
time. In absence of losses the average inversion, defined by its value
during the collapse, is greater than zero, close to $0.5,$ when we consider
the value $\frac{\chi _{0}}{\lambda _{0}}$\ $=0.2$. However, as well known,
neglecting the presence of the Kerr medium ($\frac{\chi _{0}}{\lambda _{0}}%
=0 $), the\ value of the average inversion vanishes.

Here the third-order nonlinear susceptibility represents the coupling of the 
\textit{NR} and the Kerr medium. The higher the value of $\chi $ the
stronger the coupling of the \textit{NR} with the Kerr medium, and
reversely. Considering the case $\frac{\chi _{0}}{\lambda _{0}}=0.2$ and the
system with no loss ($\kappa (t)=\delta (t)=0$)\ we observe the
collapse-revival effect (see Fig.(\ref{figInv1} a). In this figure, the
horizontal line crossing the value $I(t)=0.5$ of the average inversion is
due to the presence of nonlinearity, whose absence turns the average
inversion null. We note in this figure the amplitude of oscillation
decreasing with time, due to the presence of the Kerr medium. However, when
considering only the \textit{CPB} loss ($\kappa (t)\neq 0$, $\delta (t)=0$)
the collapse-revival effect reappears, with subsequent oscillations that
smoothly vanish (see Fig. (\ref{figInv1} b)). Contrarily, when considering
only loss in the \textit{NR} ($\kappa (t)=0$, $\delta (t)\neq 0$) these
oscillations rapidly vanish. So the spoiling effect caused by the \textit{NR}
loss dominates that coming from the \textit{CPB} loss, see Fig. (\ref%
{figInv1} c) and (\ref{figInv1} b).

%

\subsection{Off-resonant\ case: $f(t)=\protect\tau \sin (\protect\omega %
\prime )$}

\begin{figure}[h!tb]
\centering
\subfigure(a){\includegraphics[height=6cm]{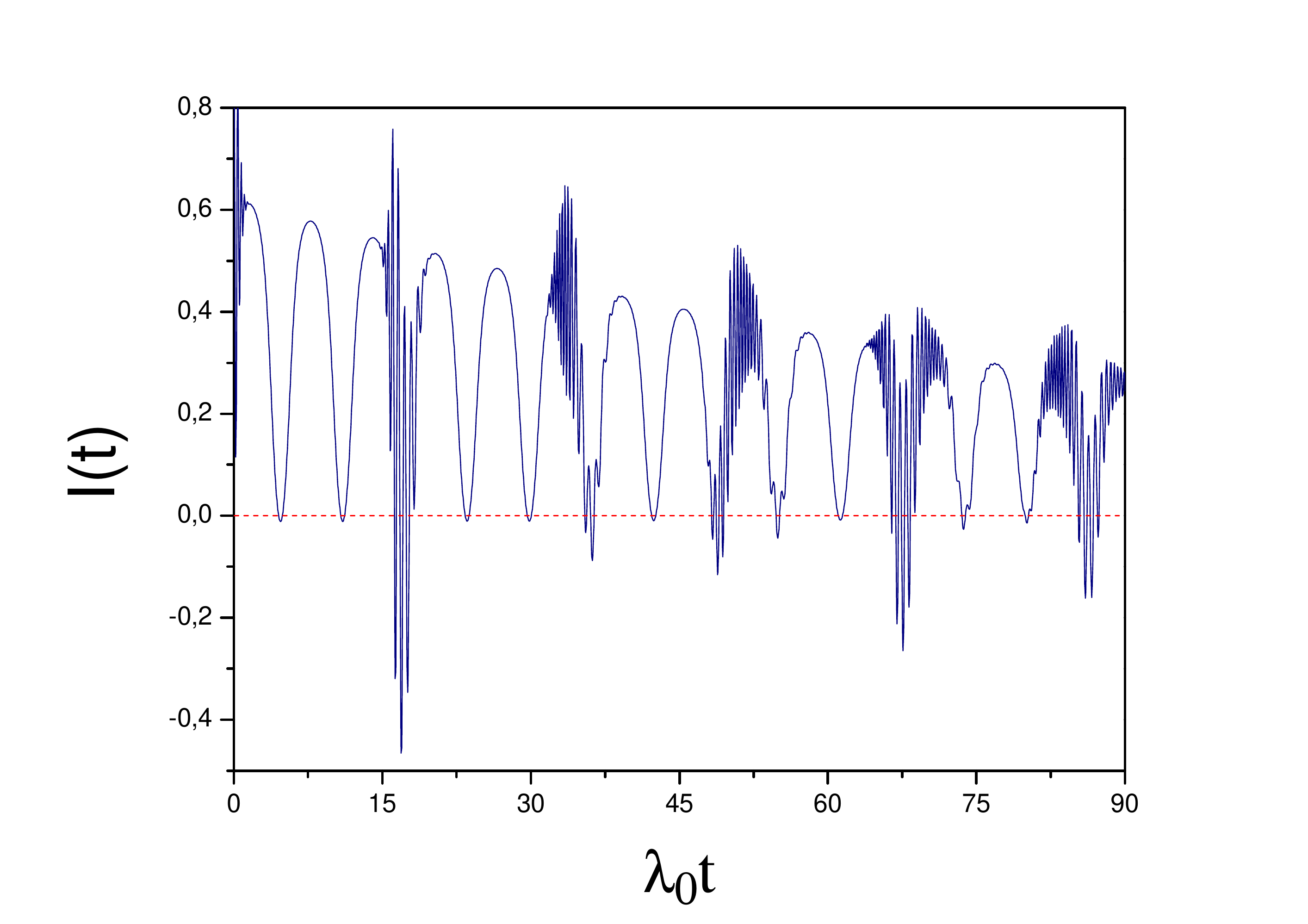} \label{fig10} } \quad 
\subfigure(b){\includegraphics[height=6cm]{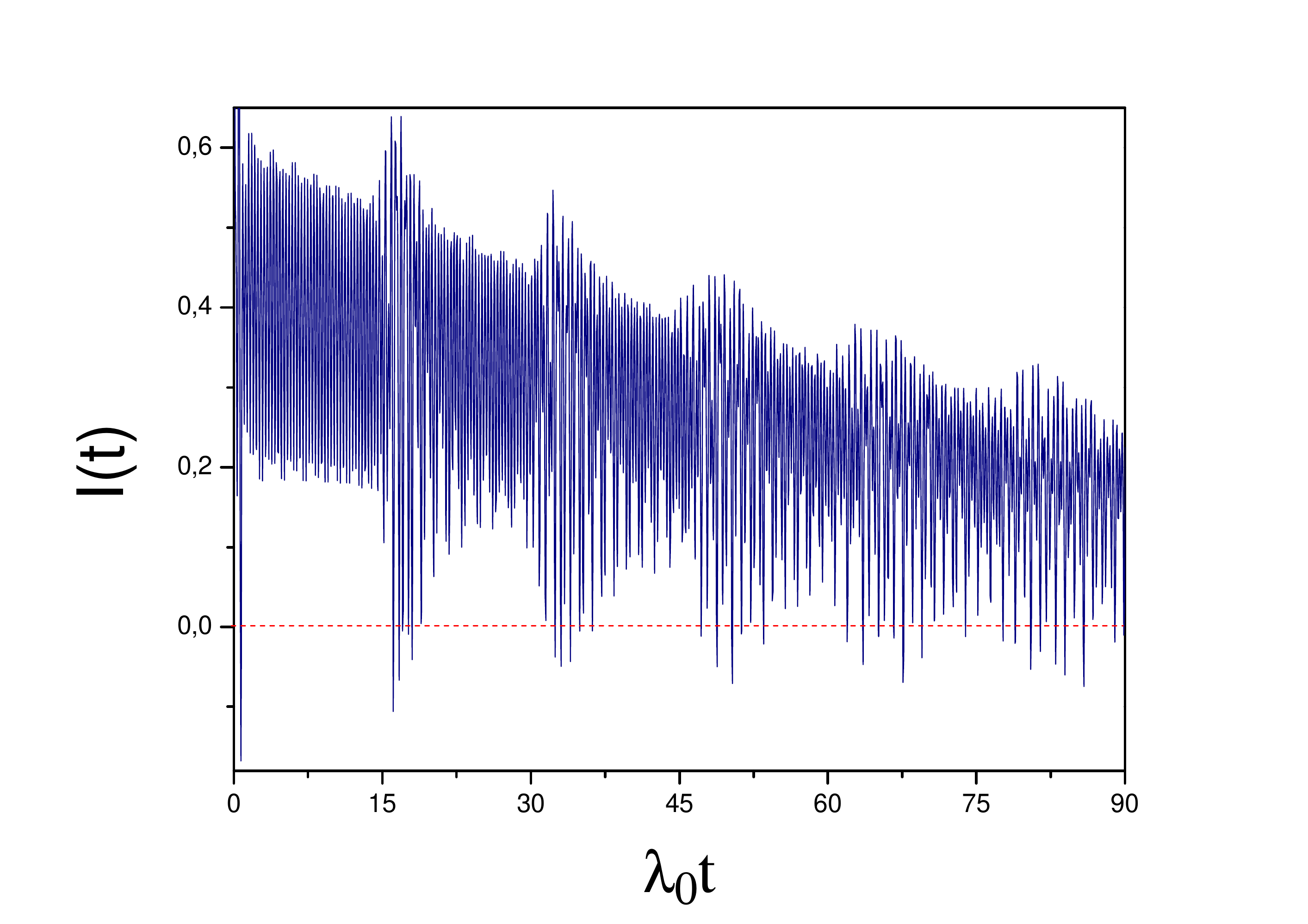} \label{fig11} } \quad 
\subfigure(c){\includegraphics[height=6cm]{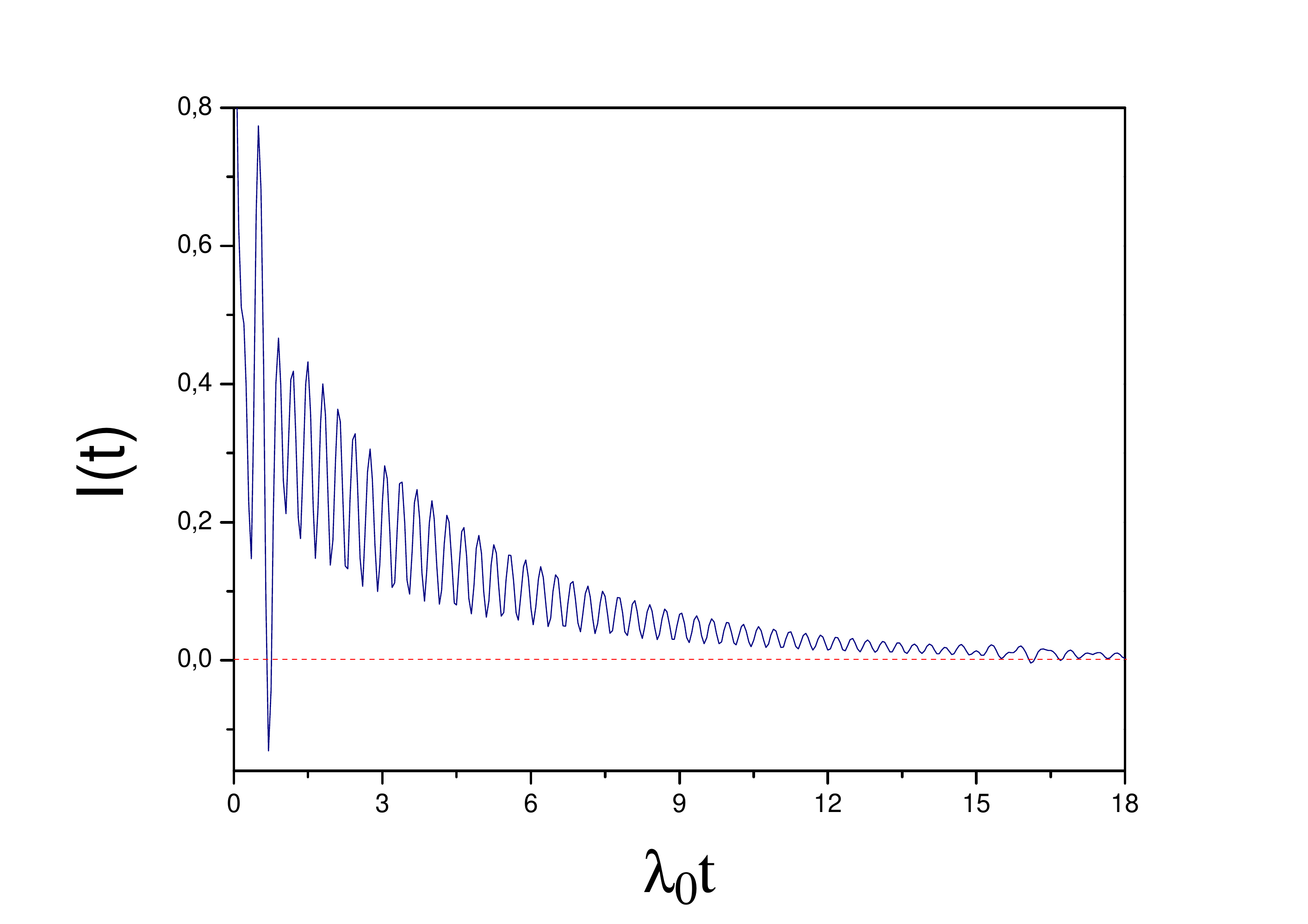} \label{fig12} } \quad 
\caption{Time evolution of the population inversion for different values of
the parameters $\protect\kappa (t)$ and $\protect\delta (t)$ for: $%
\left\langle n\right\rangle =25$, $\protect\omega _{0}/\protect\lambda _{0}=%
\protect\omega _{c}/\protect\lambda _{0}=20k$, $\protect\varepsilon /\protect%
\lambda _{0}=0.001$; $\protect\tau /\protect\lambda _{0}=10$; $f(t)=\protect%
\tau \sin (\protect\omega \prime t)$; $\protect\chi _{0}/\protect\lambda %
_{0}=0.2;$ (a) $\protect\kappa /\protect\lambda _{0}=0.01$; $\protect\delta /%
\protect\lambda _{0}=0.0$ and $\protect\omega \prime /\protect\lambda %
_{0}=1; $ (b) $\protect\kappa /\protect\lambda _{0}=0.01$; $\protect\delta /%
\protect\lambda _{0}=0.0$ and $\protect\omega \prime /\protect\lambda %
_{0}=20;$ (c) $\protect\kappa /\protect\lambda _{0}=0.0$; $\protect\delta /%
\protect\lambda _{0}=0.01$ and $\protect\omega \prime /\protect\lambda %
_{0}=20.$}
\label{figInv2}
\end{figure}

When the system is non resonant, with $\frac{\tau }{\lambda _{0}}=10$ and $%
\frac{\omega \prime }{\lambda _{0}}=1,$ and assuming only \textit{CPB} loss (%
$\kappa (t)\neq 0,$\ $\delta (t)=0$), we observe periodical behavior of the
population inversion, its amplitude decaying over time - see Fig.(\ref%
{figInv2} a); in this case the collapse-revival effect is not observed.
However, this behavior is modified when the loss comes from the \textit{NR}:
in this case the inversion and the oscillations are destroyed. Considering $%
\omega ^{\prime }$\ increasing and only the \textit{CPB} loss - see Fig.(\ref%
{figInv2} b), the system continues displaying the population inversion and
oscillations, the latter exhibiting small periods, an effect not shown when
the loss comes only from the \textit{NR} - see Fig.(\ref{figInv2} c).

\section{Time Evolution of the Entropy}

Nowdays some devices are based on quantum mechanical phenomena, and this
holds also for information transmission. For example, in optical
communication a polarized photon can carry information. Now, the entropy of
entanglement is defined for pure states as the von Neumann entropy of one of
the reduced states, e.g., the \textit{NR} entropy,\emph{\ } 
\begin{equation}
S_{NR}(t)=-\left\{ \pi _{NR}^{+}(t)\ln \pi _{NR}^{+}(t)+\pi _{NR}^{-}(t)\ln
\pi _{NR}^{-}(t)\right\} ,  \label{ent}
\end{equation}%
where $\pi _{NR}^{\pm }(t)=\frac{1}{2}(\left\langle C|C\right\rangle
+\left\langle S|S\right\rangle \pm \frac{1}{2}[(\left\langle
C|C\right\rangle -\left\langle S|S\right\rangle )^{2}+4\left\vert
\left\langle C|S\right\rangle \right\vert ^{2}]^{1/2})$ \ with $\left\langle
C|C\right\rangle =\sum_{n=0}^{\infty }\left\vert C_{1,n}(t)\right\vert ^{2}$%
, $\left\langle S|S\right\rangle =\sum_{n=0}^{\infty }\left\vert
C_{0,n+1}(t)\right\vert ^{2},$ and $\left\langle C|S\right\rangle
=\left\langle S|C\right\rangle ^{\ast }=\sum_{n=0}^{\infty }C_{1,n+1}^{\ast
}(t)C_{0,n+1}(t).$

In our calculations using the Eq. (\ref{ent}) we will assume the \textit{NR}
frequency varying in the form $\omega \left( t\right) =\omega _{0}+f(t)$.
The third order nonlinear susceptibility is modulated in the form $\chi
(t)=\chi _{0}+\varepsilon f(t)$ and we also assume the initial \textit{NR}
in a coherent state with the mean number of photon$\ \bar{n}=25$ and $\frac{%
\omega _{c}}{\lambda _{0}}=\frac{\omega _{0}}{\lambda _{0}}=20k$. Next we
will consider the time evolution of the entropy for different values of the
decay coefficients $\kappa (t)$ and $\delta (t)$.

\subsection{Resonant case: $f(t)=0$}

\begin{figure}[h!tb]
\centering
\subfigure(a){\includegraphics[height=6cm]{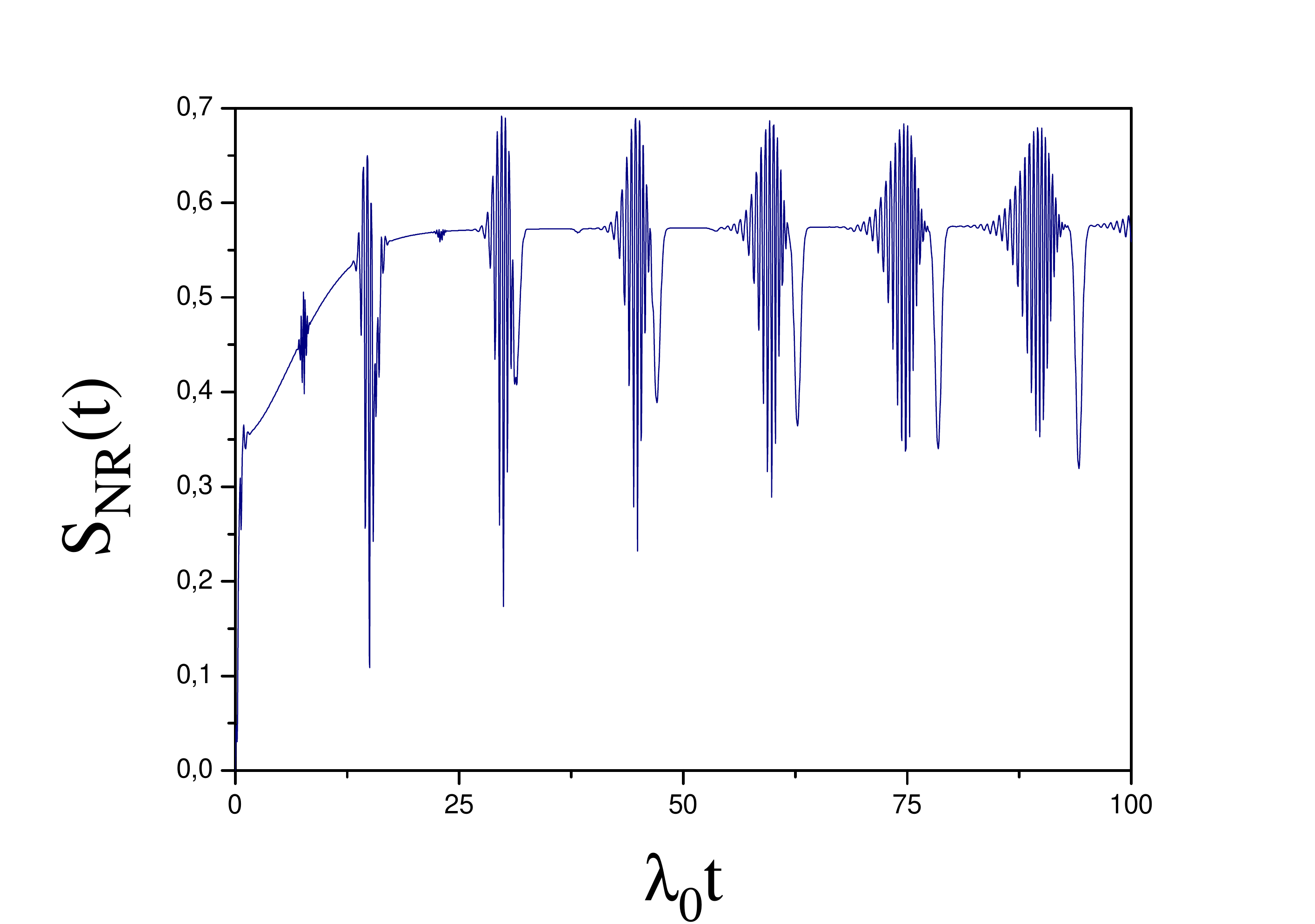} \label{s1a} } \quad 
\subfigure(b){\includegraphics[height=6cm]{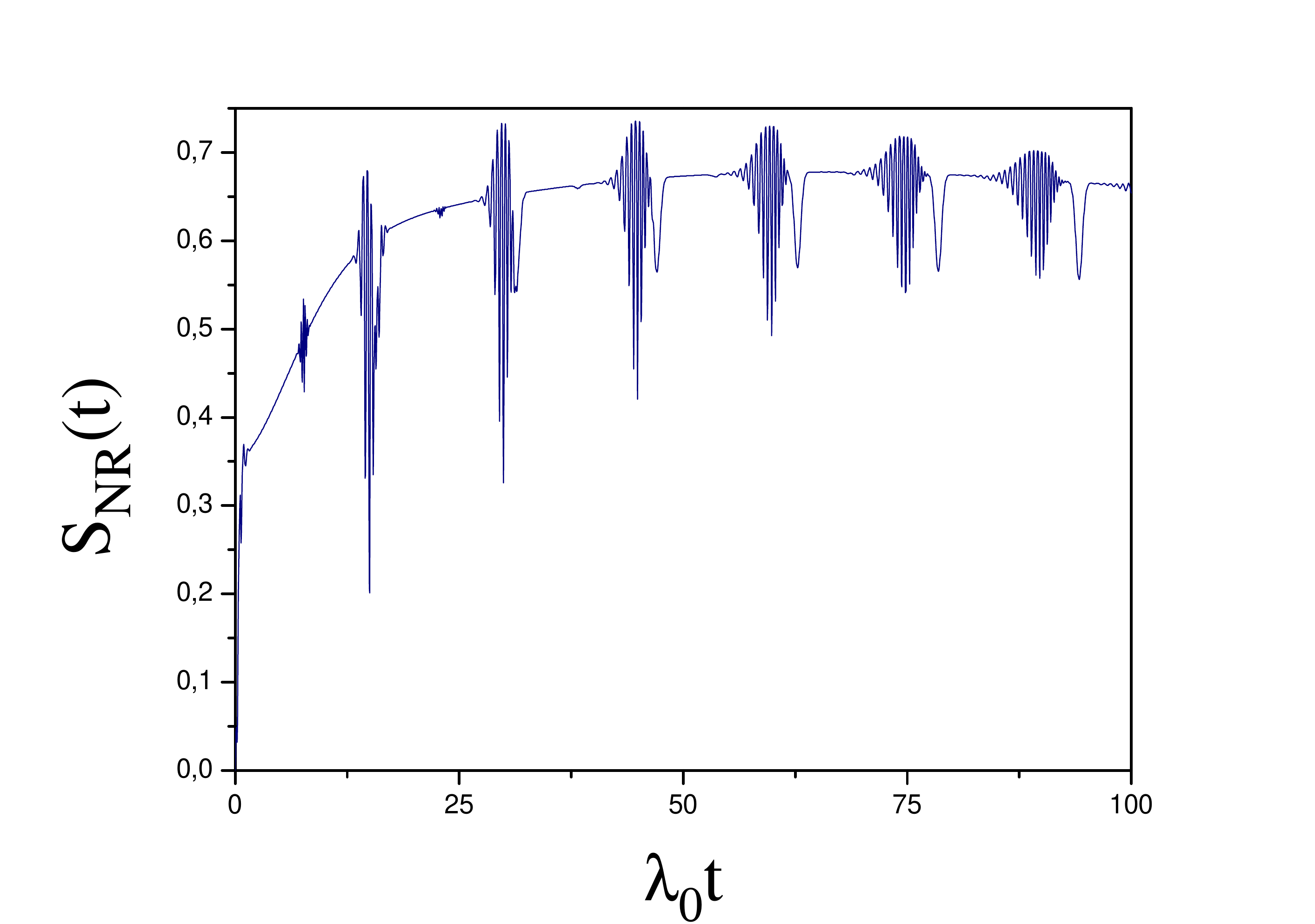} \label{s1b} }\quad 
\subfigure(b){\includegraphics[height=6cm]{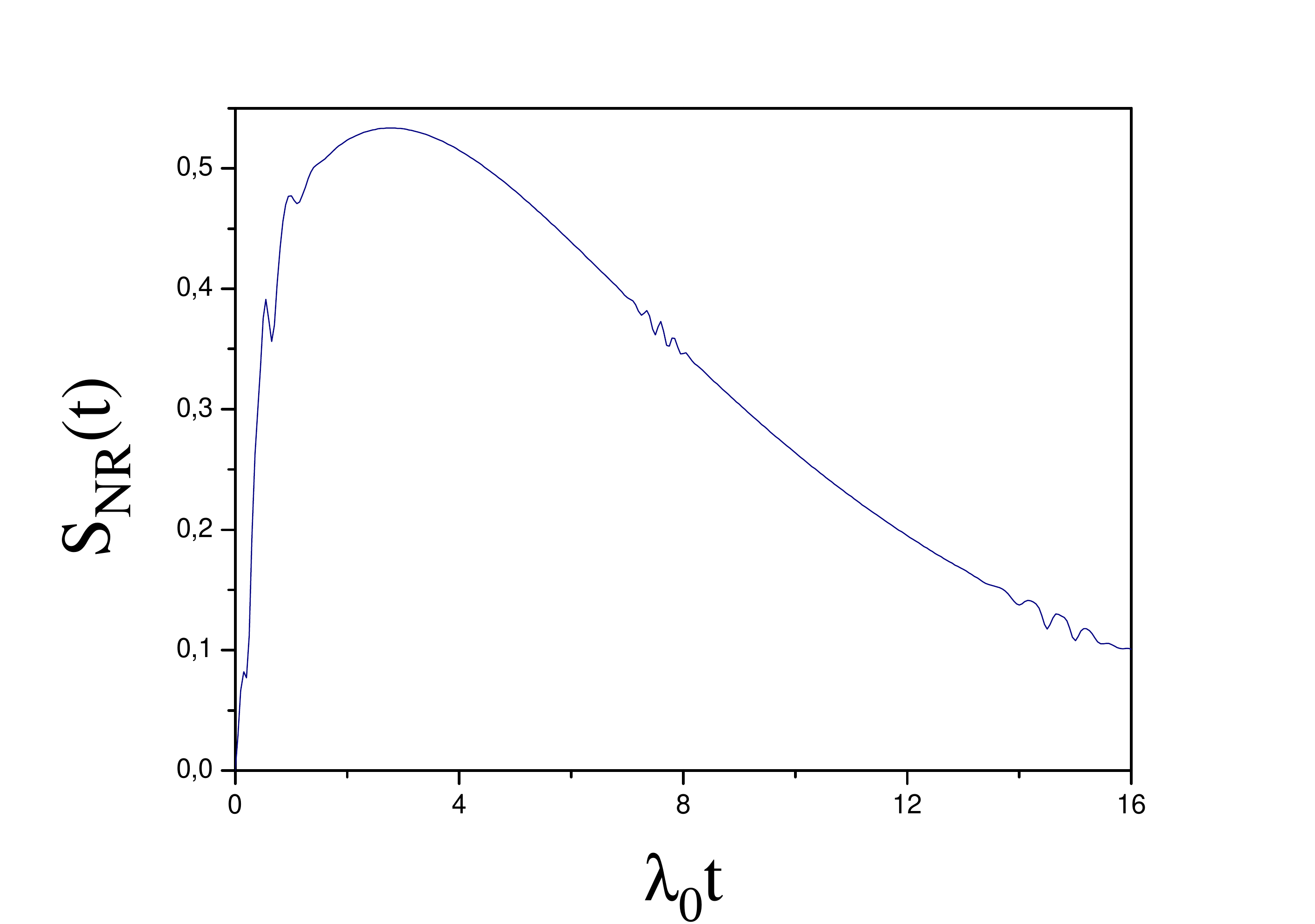} \label{s1c} }
\caption{Time evolution of the entropy for different values of the
parameters $\protect\kappa (t)$ and $\protect\delta (t)$ for: $\left\langle
n\right\rangle =25$, $\protect\omega _{c}/\protect\lambda _{0}=\protect%
\omega _{0}/\protect\lambda _{0}=20k$, $\protect\chi _{0}/\protect\lambda %
_{0}=0.2,~$ $f(t)=0$. (a) $\protect\kappa /\protect\lambda _{0}=0,0\protect%
\lambda _{0}$\ and $\protect\delta /\protect\lambda _{0}=0.0$; (b) $\protect%
\kappa /\protect\lambda _{0}=0.01$\ and $\protect\delta /\protect\lambda %
_{0}=0.0$; (c) $\protect\kappa /\protect\lambda _{0}=0.0$\ and $\protect%
\delta /\protect\lambda _{0}=0.01$.}
\label{s1}
\end{figure}

We will use the values of parameters in Fig. (\ref{s1}). In an ideal system
the parameters $\kappa (t)$ and $\delta (t)$ are null, as assumed in Fig. (%
\ref{s1} a): in this figure the maximum of the \textit{NR }entropy is close
to $\ln 2$; after the start of the interaction the \textit{NR} entropy
gradually tends to its minimum, then returns to its maximum and remains
oscillating regularly.

For small values of decay in the \textit{CPB}, as $\frac{\kappa }{\lambda
_{0}}=0.01,$ and ideal \textit{NR} $(\delta =0)$ for the case $\frac{\chi
_{0}}{\lambda _{0}}=0.2,$\ the maximum value of the entropy shows no
significative changes for small times - see Fig. (\ref{s1} b). For larger
values of time the amplitude of the entropy oscillations decreases,
maintaining the periodicity. If instead we add a small loss only in the 
\textit{NR}, say $\frac{\delta }{\lambda _{0}}=0.01,$ the entropy
oscillations vanishes rapidly (cf. Fig. (\ref{s1} c)), showing the entropy
being more sensitive to the loss in the \textit{NR} than that in the \textit{%
CPB}. For larger value of $\delta (t)$ and $\kappa (t)$ the entropy movies
rapidly to zero, as expected, due to the passage of both subsystems to their
respective ground states.

\subsection{Off-resonant\ case: $f(t)=\protect\tau \sin (\protect\omega %
\prime )$}

Let us now consider the variation in the detuning parameter, where $\tau $
and $\omega \prime $ are parameters that modulates the \textit{NR}
frequency. Our discussion is limited to the condition $\tau \ll \omega _{c}$%
, $\omega _{0}$ and also assuming that $\omega \prime $\ is small to avoid
interaction of the \textit{CPB} with other modes of the \textit{NR}. We have
chosen various values $\tau $ of amplitude modulations to verify the
entanglement properties between the \textit{CPB} and \textit{NR}. We also
use various values $\omega \prime $ of frequency modulation to see its
influence upon the \textit{CPB}-\textit{NR} entanglement - see Figs. (\ref%
{s3}). 
\begin{figure}[h!tb]
\centering
\subfigure(a){\includegraphics[height=6cm]{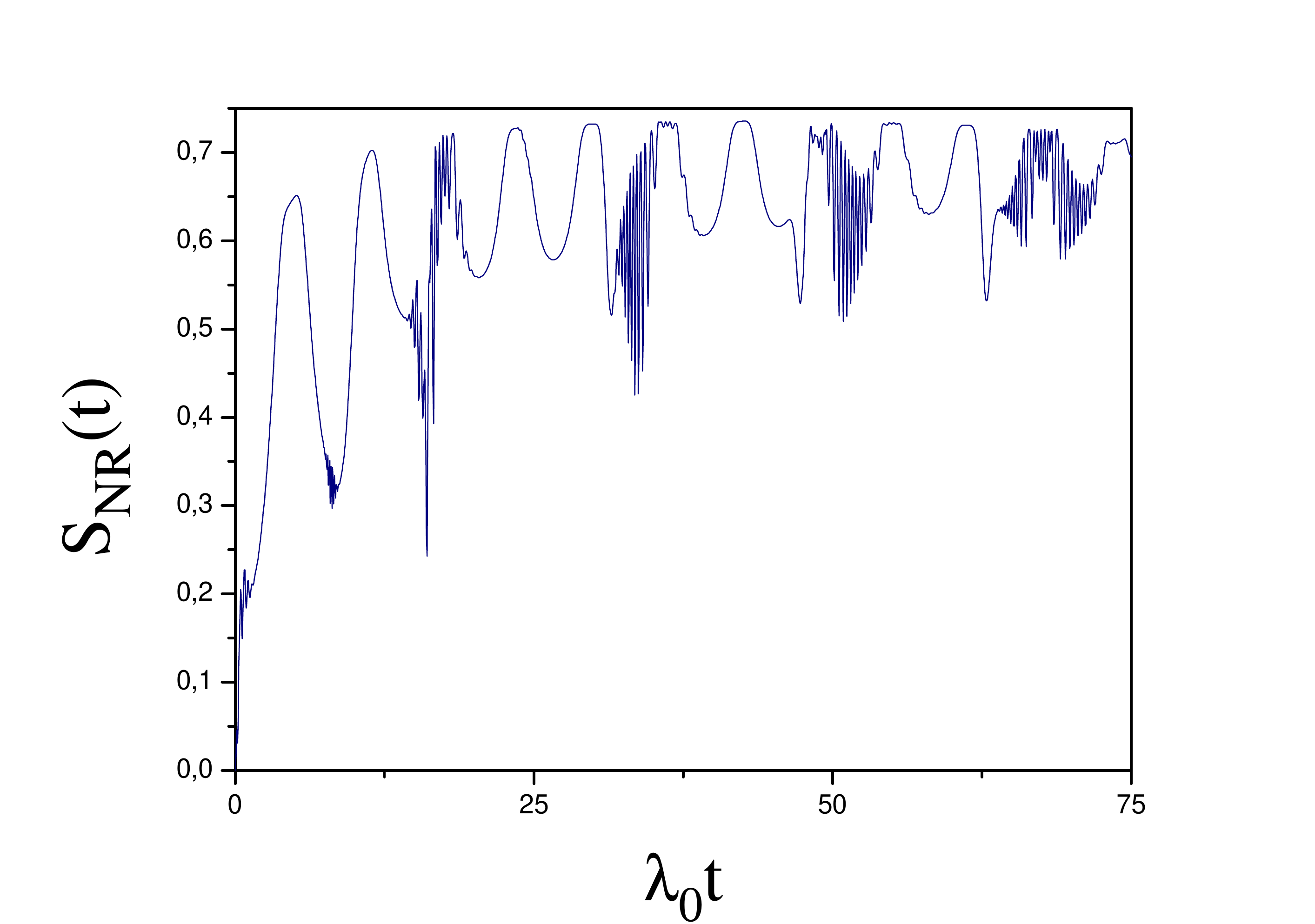} \label{i1a} } \quad 
\subfigure(b){\includegraphics[height=6cm]{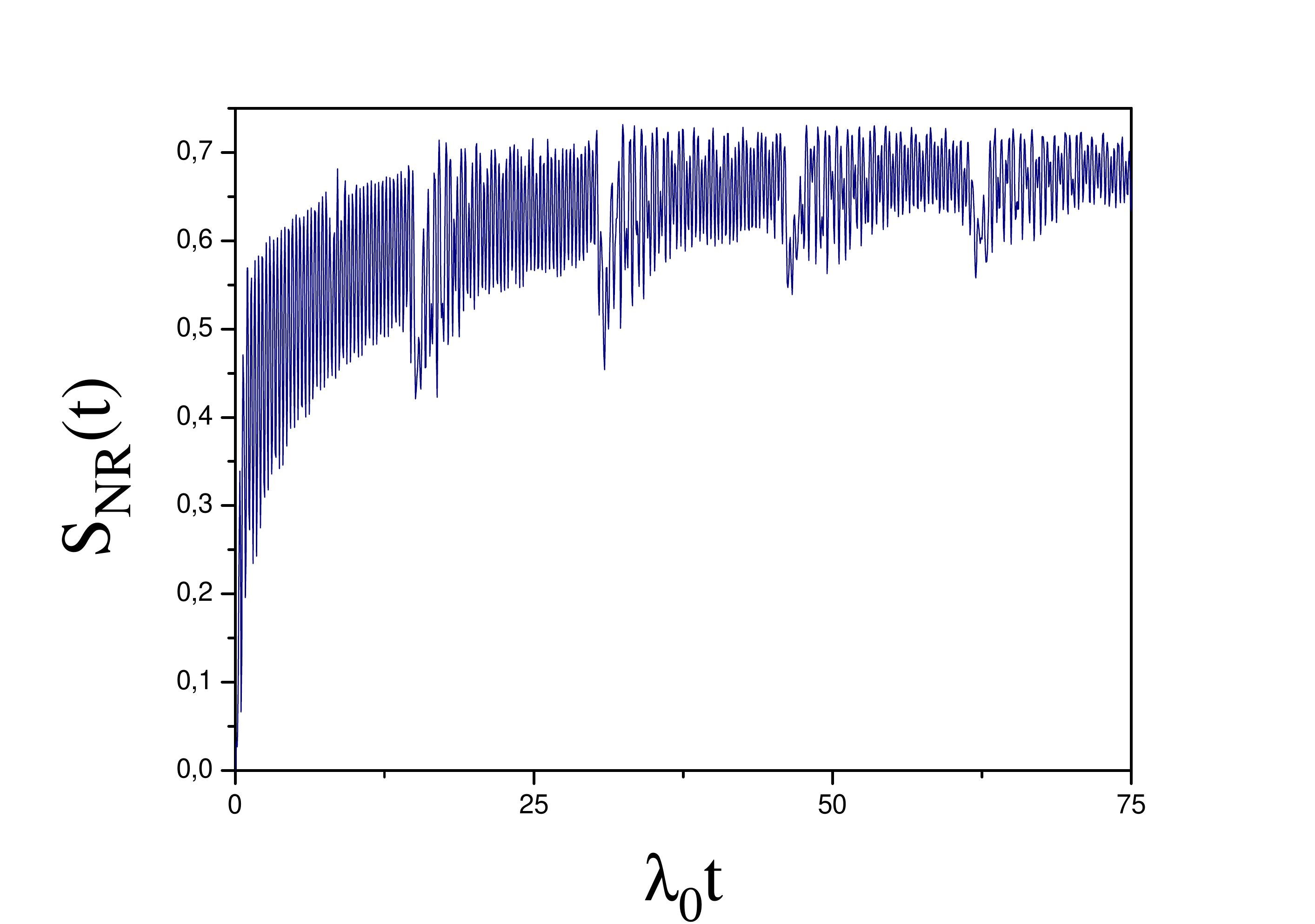} \label{i1b} } \quad 
\subfigure(c){\includegraphics[height=6cm]{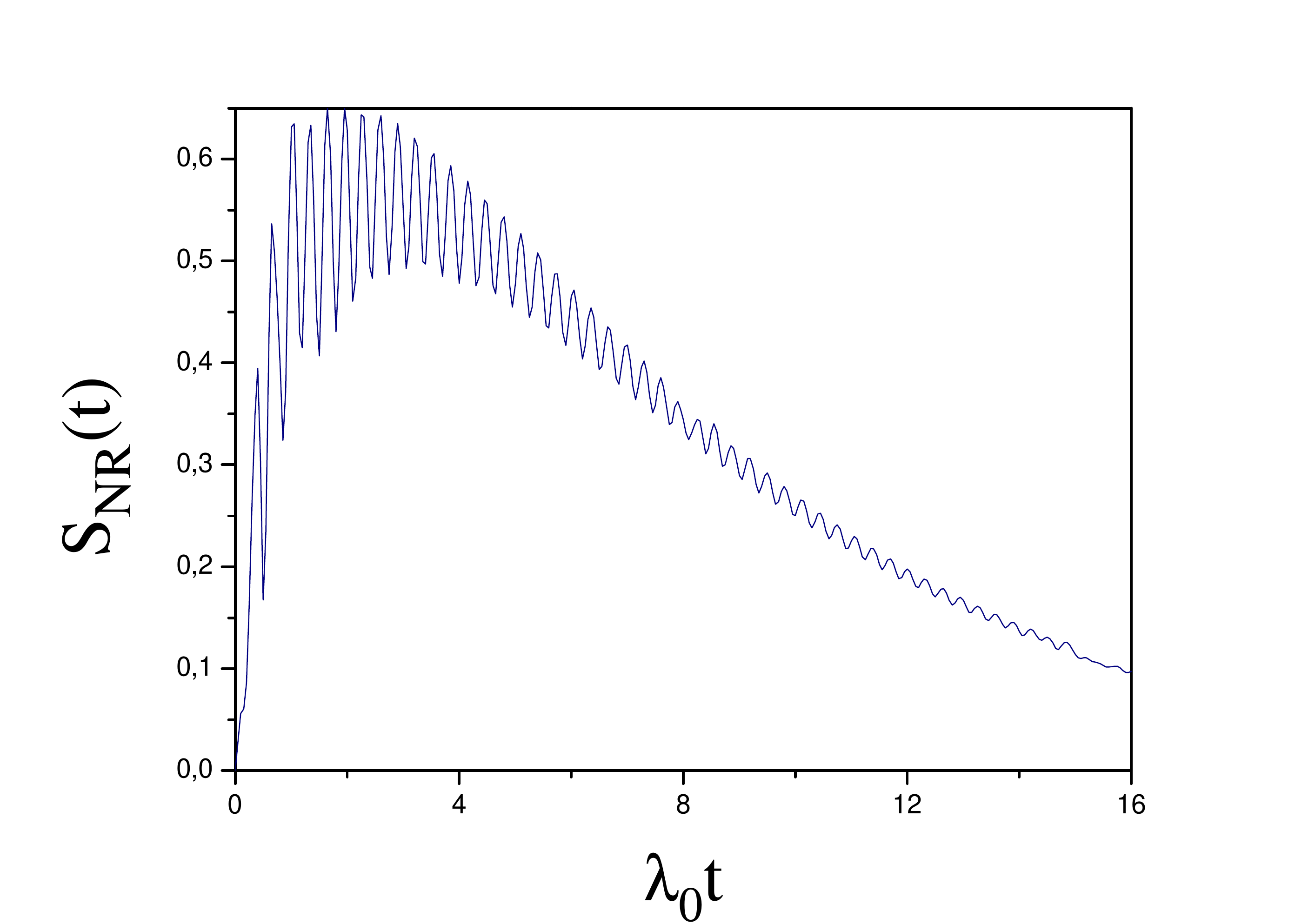} \label{i1c} }
\caption{Time evolution of the entropy for different values of the
parameters $\protect\kappa (t)$ and $\protect\delta (t)$ for: $\left\langle
n\right\rangle =25$, $\protect\omega _{c}/\protect\lambda _{0}=\protect%
\omega _{0}/\protect\lambda _{0}=20k$, $\protect\chi _{0}/\protect\lambda %
_{0}=0.2,$ $\protect\tau /\protect\lambda _{0}=10,$ and $\protect\varepsilon %
/\protect\lambda _{0}=0.001.$ (a) $\protect\kappa /\protect\lambda _{0}=0.01$%
, $\protect\delta /\protect\lambda _{0}=0.0,~$ and $\protect\omega \prime /%
\protect\lambda _{0}=1;$ (b) $\protect\kappa /\protect\lambda _{0}=0.01$, $%
\protect\delta /\protect\lambda _{0}=0.0,$ and $\protect\omega \prime /%
\protect\lambda _{0}=20;$ (c) $\protect\kappa /\protect\lambda _{0}=0.0$, $%
\protect\delta /\protect\lambda _{0}=0.01,$ and $\protect\omega \prime /%
\protect\lambda _{0}=20.$}
\label{s3}
\end{figure}
Comparing the Fig. (\ref{s1} a) with the Fig. (\ref{s3} a), we note that the
entropy loses its periodical oscillations, and when $\omega \prime $\
increases this behavior becomes more evident - see Fig. (\ref{s3} b). Now,
comparing the entropy in Fig. (\ref{s1} c) with that in Fig. (\ref{s3} c),
keeping the same parameters, we see the entropy going to zero in both cases,
but the Fig. (\ref{s3} c) displays an interesting effect: even in presence
losses, the maximum value of the entropy grows, reaching the value $\ln 2$,
after which it oscillates downward. This effect came from the sinusoidal
modulation of the \textit{NR} frequency

\section{Conclusion}

In the present work we have considered the interaction of a \textit{CPB} and
an \textit{NR} in the presence of a Kerr medium and losses affecting both
subsystems. Concerning the influence of the loss affecting the entropy of
both subsystems and the population inversion of the \textit{CPB} we have
observed the dominant role played by the \textit{NR} upon that played by the 
\textit{CPB}. The dissipation causes deterioration of the \textit{CPB}
excited level whereas convenient modulations favors the control of certain
properties of the system. It was also shown that certain choice of the
time-dependent frequency\ makes higher the maximum value of the entropy,
even in the presence of dissipation (see Figs. (\ref{s1} c) and Fig. (\ref%
{s3} c) ) the same occurring for the \textit{CPB} population inversion - see
Figs. (\ref{figInv1} c) and Fig. (\ref{figInv2} c). Concerning the entropy,
this result is very important for information transmission since the
transmission of maximum information through a quantum channel is exactly the
von Neumann entropy \cite{q1,q4}. These effects are very sensitive to
detuning ($\approx 0.04\%$) and disappear in resonant regime. The results
suggest that it is possible to perform a dynamic control of certain
properties of this system, via convenient manipulation of the parameters. We
hope that these results might\ shed light in this scenario, furnishing new
insights for researchers in this area.

\section{Acknowledgments}

We thank the CNPq and FAPEG for the partial supports.


\begin{thebibliography}{99}
\bibitem{15c} Y. Nakamura, Y. A. Pashkin, and J. Tsai, \textit{Coherent
control of macroscopic quantum states in a single-cooper-pair box}, Nature 
\textbf{398}, 798 (1999).

\bibitem{16c} A. Wallraff, D. Schuster, A. Blais, L. Frunzio, R.-S. Huang,
J. Majer, S. Kumar, S. Girvin, and R. Schoelkopf, \textit{Strong coupling of
a single photon to a superconducting qubit using circuit quantum
electrodynamics}, Nature \textbf{431}, 162 (2004).

\bibitem{24c} M. Zhang, J. Zou, and B. Shao, \textit{Quantum dynamics of a
single cooperpair box with a single-mode cavity field}, Int. J. Mod. Phys. B 
\textbf{16}, 4767, 2002.

\bibitem{25c} A.-S. Obada, N. Metwally, D. M. Abo-kahla, and M. abdel Aty, 
\textit{The quantum computational speed of as single cooper pair box},
Physica E \textbf{43}, 1792 (2011).

\bibitem{26c} A.-H. M. Ahmed, L. Y. Cheong, N. Zakaria, and N. Metwally, 
\textit{Dynamics of information coded in a single cooper pair box}, Int. J.
Theor. Phys. \textbf{52}, 1979 (2013).

\bibitem{31c} Z. Bian, F. Chudak, W. G. Macready, L. Clark, and F. Gaitan, 
\textit{Experimental determination of ramsey numbers}, Phys. Rev. Lett. 
\textbf{111}, 130505 (2013).

\bibitem{c3} C. Valverde and B. Baseia, \textit{Engineering information in
states of a nanomechanical resonator coupled to a Cooper pair box}, Quantum
Inf. Process. \textbf{12}, 2019 (2013).

\bibitem{c4} C. Valverde, A. T. Avelar, and B. Baseia, \textit{Quantum
communication via controlled holes in the statistical distribution of
excitations in a nanoresonator coupled to a Cooper pair box}, Chin. Phys. B 
\textbf{21}, 030308 (2012).

\bibitem{15} C. K. Law, S. Y. Zhu, and M. S. Zubairy, \textit{Modification
of a vacuum Rabi splitting via a frequency modulated cavity mode}, Phys.
Rev. A \textbf{52}, 4095 (1995).

\bibitem{15a} M. Janowicz, \textit{Evolution of wave fields and atom-field
interactions in a cavity with one oscillating mirror}, Phys. Rev. A \textbf{%
57}, 4784 (1998).

\bibitem{16} M. S. Abdalla, M. Abdel-Aty, and A. S. F. OBADA, \textit{%
Entropy and entanglement of time dependent two-mode Jaynes--Cummings model,}%
\textbf{\ }Physica A \textbf{326}, 203 (2003).

\bibitem{17} Y. P. Yang, J. P. Xu,, LI, G. X. e CHEN, H., \textit{%
Interactions of a two-level atom and a field with a time-varying frequency,}
Phys. Rev. A \textbf{69}, 053406 (2004).

\bibitem{nn3} S.N. Shevchenko, S. Ashhab and F. Nori, \textit{%
Landau--Zener--St\"{u}ckelberg interferometry},\ Physics Reports \textbf{492}%
, 1 (2010).

\bibitem{nn4} J. Q. You and F. Nori, \textit{Atomic physics and quantum
optics using superconducting circuits}, Nature \textbf{474}, 589 (2011).

\bibitem{nn5} P.D. Nation, J.R. Johansson, M.P. Blencowe, and F. Nori, 
\textit{Colloquium: Stimulating uncertainty: Amplifying the quantum vacuum
with superconducting circuits}, Rev. Mod. Phys. \textbf{84}, 1 (2012).

\bibitem{nn6} I.\thinspace M. Georgescu, S. Ashha and F. Nori, \textit{%
Quantum simulation}, Rev. Mod. Phys. \textbf{86}, 153 (2014).

\bibitem{nn2} I. Buluta, S. Ashhab and F. Nori, \textit{Natural and
artificial atoms for quantum computation}, Reports on Progress in Physics 
\textbf{74}, 104401 (2011).

\bibitem{nn7} C. P. Sun, L. F. Wei, Yu-xi Liu, and F. Nori, \textit{Quantum
transducers: Integrating transmission lines and nanomechanical resonators
via charge qubits}, Phys. Rev. A \textbf{73}, 022318 (2006).

\bibitem{nn8} Yu-xi Liu, A. Miranowicz, Y. B. Gao, J. Bajer, C. P. Sun and
F. Nori, \textit{Qubit-induced phonon blockade as a signature of quantum
behavior in nanomechanical resonators}, Phys. Rev. A \textbf{82}, 032101
(2010).

\bibitem{nn9} Fei Xue, Yu-xi Liu, C. P. Sun and F. Nori, \textit{Two-mode
squeezed states and entangled states of two mechanical resonators}, Phys.
Rev. B \textbf{76}, 064305 (2007).

\bibitem{nn10} M. Grajcar, S. Ashhab, J. R. Johansson and F. Nori, \textit{%
Lower limit on the achievable temperature in resonator-based sideband cooling%
}, Phys. Rev. B \textbf{78}, 035406 (2008).

\bibitem{nn11} J. R. Johansson, N. Lambert, I. Mahboob, H. Yamaguchi and F.
Nori, \textit{Entangled-state generation and Bell inequality violations in
nanomechanical resonators}, Phys. Rev. B \textbf{90}, 174307 (2014).

\bibitem{c1} C. Valverde, A.T. Avelar, and B. Baseia, \textit{Controlling
statistical properties of a Cooper pair box interacting with a
nanomechanical resonator}, Physica A \textbf{390}, 4045 (2011).

\bibitem{c2} C. Valverde, H. C. B. de Oliveira, A. T. Avelar, and B. Baseia, 
\textit{Controlling Excitation Inversion of a Cooper Pair Box Interacting
with a Nanomechanical Resonator}, Chin. Phys. Lett. 29, 080303 (2012).

\bibitem{d1} F. Bagarello, M. Lattuca, R. Passante, L. Rizzuto and S.
Spagnolo, \textit{Non-Hermitian Hamiltonian for a modulated Jaynes-Cummings
model with }$PT$\ \textit{symmetry}, Phys. Rev. A \textbf{91}, 042134 (2015).

\bibitem{d2} B. Peng, S. K. Ozdemir, F. Lei, F. Monifi, M. Gianfreda, G. L.
long, S. Fan, F. Nori, C. M. Bender, and L. Yang, \textit{%
Parity-timesymmetric whispering-gallery microcavities}, Nat. Phys. \textbf{10%
}, 394 (2014).

\bibitem{d3} C. M. Bender and S. Boettcher, \textit{Real spectra of
non-hermitian hamiltonian having }$PT$\ \textit{\ symmetry}, Phys. Rev.
Lett. \textbf{80}, 5243(1998).

\bibitem{d4} C. M. Bender, D. C. Brody, and H. F. Jones, \textit{Complex
extension of quantum mechanics}, Phys. Rev. Lett. \textbf{89}, 270401 (2002).

\bibitem{d5} C. M. Bender, \textit{Making sense of non-Hermitian hamiltonians%
}, Rep. Prog. Phys. \textbf{70}, 947 (2007).

\bibitem{q1} A. Peres, \textit{Quantum Theory: Concepts and Methods}, Kluwer
(1993).

\bibitem{q4} D. Petz, \textit{Properties of quantum entropy, in Quantum
Probability and Related Topics VII}, 275--297, World Sci. Publishing (1992).
\end{thebibliography}
\end{document}